\documentclass[11pt]{article}
\usepackage{latexsym,amssymb}

\newcommand{\eq}[1]{(\ref{#1})}

\newcommand{\be}{\begin{equation}}
\newcommand{\ee}{\end{equation}}
\newcommand{\bea}{\begin{eqnarray}}
\newcommand{\eea}{\end{eqnarray}}
\newcommand{\nn}{\nonumber}

\newcommand{\ba}{\begin{eqnarray}}
\newcommand{\ea}{\end{eqnarray}}

\def\Z{\mathcal{Z}}

\def\jbar{{\bar \jmath}}

\def\IP{\relax{\rm I\kern-.18em P}}

\font\cmss=cmss10 \font\cmsss=cmss10 at 7pt
\def\twomat#1#2#3#4{\left(\matrix{#1 & #2 \cr #3 & #4}\right)}
\def\inbar{\vrule height1.5ex width.4pt depth0pt}
\def\IC{\relax\,\hbox{$\inbar\kern-.3em{\rm C}$}}
\def\IG{\relax\,\hbox{$\inbar\kern-.3em{\rm G}$}}
\def\IB{\relax{\rm I\kern-.18em B}}
\def\ID{\relax{\rm I\kern-.18em D}}
\def\IL{\relax{\rm I\kern-.18em L}}
\def\IF{\relax{\rm I\kern-.18em F}}
\def\IH{\relax{\rm I\kern-.18em H}}
\def\II{\relax{\rm I\kern-.17em I}}
\def\IN{\relax{\rm I\kern-.18em N}}
\def\IP{\relax{\rm I\kern-.18em P}}
\def\IQ{\relax\,\hbox{$\inbar\kern-.3em{\rm Q}$}}
\def\bfzero{\relax\,\hbox{$\inbar\kern-.3em{\rm 0}$}}
\def\IK{\relax{\rm I\kern-.18em K}}
\def\IG{\relax\,\hbox{$\inbar\kern-.3em{\rm G}$}}
 \font\cmss=cmss10 \font\cmsss=cmss10 at 7pt
\def\IR{\relax{\rm I\kern-.18em R}}
\def\ZZ{\relax\ifmmode\mathchoice
{\hbox{\cmss Z\kern-.4em Z}}{\hbox{\cmss Z\kern-.4em Z}}
{\lower.9pt\hbox{\cmsss Z\kern-.4em Z}} {\lower1.2pt\hbox{\cmsss
Z\kern-.4em Z}}\else{\cmss Z\kern-.4em Z}\fi}
\def\bfone{\relax{\rm 1\kern-.35em 1}}

\def\IU{\relax\,\hbox{$\inbar\kern-.3em{\rm U}$}}

\def\be{\beta}

\def\part{\partial}

\def\square{{\,\lower0.9pt\vbox{\hrule \hbox{\vrule height 0.2 cm
\hskip 0.2 cm \vrule height 0.2 cm}\hrule}\,}}

\def\bfone{\relax{\rm 1\kern-.35em 1}}
\font\cmss=cmss10 \font\cmsss=cmss10 at 7pt
\def\tilde{\widetilde}
\def\bar{\overline}

\def\hat{\widehat}

\def\Coe#1.#2.{\frac{#1}{ #2}}

\def\coe#1.#2.{\relax{\textstyle {#1 \over #2}}\displaystyle}

\def\to{\rightarrow}
\def\notin{\hbox{{$\in$}\kern-.51em\hbox{/}}}

\def\IE{\relax{{\rm I\kern-.18em E}}}

\def\IGam{\relax{{\rm I}\kern-.18em \Gamma}}

\def\inbar{\vrule height1.5ex width.4pt depth0pt}
\def\bfzero{\relax{\rm I\kern-.18em 0}}
\def\bfone{\relax{\rm 1\kern-.35em 1}}
\def\twomat#1#2#3#4{\left(\begin{array}{cc}
\end{array}
\right)}

\DeclareFontFamily{U}{rsf}{} \DeclareFontShape{U}{rsf}{m}{n}{
  <5> <6> rsfs5 <7> <8> <9> rsfs7 <10-> rsfs10}{}
\DeclareMathAlphabet\Scr{U}{rsf}{m}{n}
\begin{document}

  \begin{center}{\LARGE \bf  First Order Description of $D=4$\\ static Black Holes\\\vskip 3mm and the Hamilton--Jacobi
   equation} \vskip 1.5cm {L. Andrianopoli$^{*}$,  R. D'Auria$^*$, E.
Orazi$^{*}$ and M. Trigiante$^* $}   \end{center}
 \vskip 3mm
\noindent {\small $^*$ Dipartimento di Fisica,
  Politecnico di Torino, Corso Duca degli Abruzzi 24, I-10129
  Turin, Italy and Istituto Nazionale di Fisica Nucleare (INFN)
  Sezione di Torino, Italy
 }
 \footnote{\noindent\texttt{laura.andrianopoli@polito.it};

 \texttt{riccardo.dauria@polito.it};

  \texttt{emanuele.orazi@polito.it};

   \texttt{mario.trigiante@polito.it}.}

\vfill
\vskip 1,5 cm

\begin{abstract}
In this note we discuss the application of the Hamilton--Jacobi
formalism to the first order description of four dimensional
spherically symmetric and static black holes. In particular we show
that the prepotential characterizing the flow coincides with the
Hamilton principal function associated with  the one-dimensional
effective Lagrangian. This implies that the prepotential can always
be defined, at least locally in the radial variable and in the
moduli space, both in the extremal and non-extremal case and allows
us to conclude that it is duality invariant. We also give, in this
framework, a general definition of the ``Weinhold metric'' in terms
of which a necessary condition for the existence of
multiple attractors is given. The Hamilton--Jacobi formalism can be applied
both to the restricted phase space where the electromagnetic
potentials have been integrated out as well as in the case where the
electromagnetic potentials are dualized to scalar fields using the
so-called three-dimensional Euclidean approach.  We give some
examples of application of the formalism, both for the BPS and the
non-BPS black holes.
\end{abstract}

 \vfill\eject
\section{Introduction} \label{intro}
Recently considerable attention has been devoted to the study of
black-hole solutions in supergravity theories, especially in
connection with the attractor mechanism
\cite{attractor,attractor2} (see also \cite{review} and references
therein). In this respect,
 of particular relevance is the issue of describing the
 spatial evolution of the scalar fields and the metric in terms of a
first order dynamical system of equations written in terms of a
prepotential, also called \emph{fake superpotential}
\cite{Ceresole:2007wx,Andrianopoli:2007gt,LopesCardoso:2007ky,Perz:2008kh}.
In the case of four dimensional spherically symmetric  black holes,
if we collectively denote the metric and scalar degrees of freedom
characterizing the solution by $q^i$ and the radial variable by
$\tau$, the issue has been of whether it is possible to define a
prepotential $\mathcal{W}(q)$, function of $q^i$ and of the
quantized electric-magnetic charges, such that the radial evolution
of $q^i$ is solution to a system of equations of the following form:
\begin{eqnarray}
\dot{q}^i&=&G^{ij}\,\frac{\partial \mathcal{W}}{\partial
q^j}\,,\label{dyn}
\end{eqnarray}
$G_{ij}(q)$ being a suitable non-degenerate metric. Equations
(\ref{dyn}) are particularly suitable for studying the attractor
behavior \cite{attractor,attractor2} of four dimensional black-hole
solutions \cite{marrani,marrani2} as well as higher dimensional
black-brane solutions \cite{intersecting}. The reason behind the
interest in such a prepotential $\mathcal{W}$ goes beyond the simple
convenience in writing the black-hole equations in a first order
form. This function seems to have a deeper physical meaning also
related to the description of the radial evolution of the black-hole
fields in terms of a dual RG flow.
\par So far such a description has been found for specific extremal
and non-extremal static four dimensional black-hole solutions and
the following questions have been left unanswered:
\begin{itemize}
\item{For which solutions does the prepotential $\mathcal{W}$ exist?}
\item{Is $\mathcal{W}$ invariant under the global symmetries (dualities) of
the four dimensional theory?}
\item{In ref. \cite{Andrianopoli:2007gt} it was found for the extremal case that $\mathcal{W}$
 is a c-function, that is always monotonic in $\tau$. Can this
property be extended, whenever $\mathcal{W}$ exists, also to the
non-extremal case?}
\end{itemize}
It is a well established fact that the problem of finding a first
order description of the form (\ref{dyn}) can be recast, for static
solutions, into a Hamilton--Jacobi problem
 in which the $\mathcal{W}$
prepotential is the Hamilton's characteristic function such that:
\begin{eqnarray}
H(q,\partial_q \mathcal{W})&=&c^2\,.
\end{eqnarray}
The Hamilton--Jacobi approach to the first order description of
supergravity solutions has been applied to the study of  RG flows
``dual'' to domain walls and cosmological solutions \cite{de
Boer:1999xf,Verlinde:1999xm,Fukuma:2002sb,Skenderis:2006rr} and to
extremal black holes \cite{Hotta:2009bm}. It has also been applied
to the general description of extended supergravity solutions in
various dimensions \cite{Janssen:2007rc}.
\par In this note we
restrict ourselves to static, spherically symmetric black holes and
wish to point out some important consequences of the well known
theory of the Hamilton--Jacobi equation which have never been clearly
stated in the black-hole literature so far.\par We first show that
the prepotential,  being identified with Hamilton's characteristic
function, can always be given (in both the extremal ($c=0$) and
non-extremal ($c\neq 0$) case) in a local form, in terms of the
integral of the Lagrangian along
 a trajectory which minimizes the action (characteristic trajectory).
 We also give, in this framework, a general
definition of the ``Weinhold metric'' \cite{attractor2} in terms of
which to define a necessary condition \cite{moore} for the existence
of multiple attractors in the extremal case ($c=0$).
%
%
Actually, the expression we find for the prepotential, eq. \eq{HJ1}
below, generalizes the one conjectured in
\cite{Andrianopoli:2007gt}.
\par
 Of particular interest to us is the issue of whether $\mathcal{W}(q)$ is
 duality invariant. If $\mathcal{W}$ is to be associated with some fundamental physical
 property of the black hole, just like the entropy, it ought to be
 duality invariant, namely independent on the particular
 description of its degrees of freedom. The duality invariance of
 $\mathcal{W}$ is guaranteed by the fact that it can be expressed as the
 integral of the duality invariant Lagrangian on a characteristic
 trajectory.\par The paper is organized as follows: in Section
 \ref{sec1} we briefly review the Hamiltonian description of four
 dimensional spherically symmetric and static black holes. In Section \ref{sec2}
we introduce  the Hamilton--Jacobi problem in $D=4$ and the
definition of the prepotential $\mathcal{W}$. In Section
\ref{sec25} we write a local form for the solution to the
Hamilton--Jacobi equation and discuss some of its properties. In
Section \ref{sec3} we prove duality invariance of $\mathcal{W}$.
Finally in Section \ref{sec4} we generalize the Hamiltonian
problem by extending the phase space to include the quantized
charges and their conjugate variables, namely the
electric-magnetic potentials. This extended formulation has a
natural setting in the $D=3$ Euclidean theory arising from time
reduction of the four dimensional one. We write the
Hamilton--Jacobi equation in this enlarged phase space and express
its solution $\mathcal{W}_3$ in terms of the four dimensional
prepotential $\mathcal{W}$. We discuss cases of interest in which
a globally defined solution $\mathcal{W}_3$, and correspondingly
$\mathcal{W}$, of the Hamilton--Jacobi equation exists.

\section{Review of Static $D=4$ Black Holes}
\label{sec1} Let us recall the main facts about the description of a
static black hole in $D=4$ as solution of a Hamiltonian system. We
start from the four dimensional bosonic action of a generic
supergravity theory, describing $n$ scalar fields $\phi=(\phi^r)$
coupled to $n_v$ vector fields $A^\Lambda_\mu$:
\begin{eqnarray}
S_4&=&\int
d^4x\,\sqrt{-g}\left[-\frac{1}{2}\,R+\frac{1}{2}\,G_{rs}(\phi)\,\partial_\mu
\phi^r\,\partial^\mu
\phi^s+I_{\Lambda\Sigma}(\phi)\,F_{\mu\nu}^{\Lambda}\,
F^{\Sigma\,\mu\nu}+\right.\nonumber\\&&+\left.\frac{1}{2\,\sqrt{-g}}\,\epsilon^{\mu\nu\rho\sigma}\,R_{\Lambda\Sigma}(\phi)\,F_{\mu\nu}^{\Lambda}
\,F_{\rho\sigma}^{\Sigma}\right]\,,\label{S4}
\end{eqnarray}
where the gauge field-strength 2-form is defined as $F^\Lambda = d
A^\Lambda$ and $I_{\Lambda\Sigma},\,R_{\Lambda\Sigma}$ are the imaginary and real part of the
complex kinetic matrix $\mathcal{N}_{\Lambda\Sigma}(\phi)$, with the
convention that $I_{\Lambda\Sigma}<0$. The general Ansatz for a
spherically symmetric static black hole reads
\cite{breit,attractor2,review}:
\begin{eqnarray}
ds_4^2&=&g_{\mu\nu}\,dx^\mu\,dx^\nu=e^{2\,U}\,dt^2-e^{-2\,U}\,\left(\frac{c^4}{\sinh^4(c\,\tau)}\,d\tau^2+\frac{c^2}{\sinh^2(c\,\tau)}\,d\Omega^2\right)\,,\nonumber\\
F^\Lambda&=&m^\Lambda\,\sin(\theta)\,d\theta\wedge
d\varphi+e^{2\,U}\,\ell^\Lambda(\phi)\,dt\wedge d\tau\,,
\end{eqnarray}
where
$\ell^\Lambda(\phi)=I^{-1\,\Lambda\Sigma}\,(e_\Sigma-R_{\Sigma\Gamma}\,m^\Gamma)$,
$e_\Lambda,\,m_\Sigma$ being the quantized electric and magnetic
charges. The constant $c$ is the extremality parameter which is
related to the temperature and entropy of the black hole through
$c=2\,S\,T$ and identifies the inner and outer horizons
corresponding to $r_\pm=r_0\pm c$. The dependence of the evolution
parameter $\tau $ on the radial variable $r$ is implicitly given by
the equation $\frac{c^2}{\sinh^2(c\,\tau)}=(r-r_0)^2-c^2$, which in
the extremal case $c\to 0$ can be written as $\tau =-1/(r -r_0)$.

It is well known that the equations of motion obtained from the
above Ansatz read:
\bea
\frac{d^2 U}{d\tau^2}  &\equiv& \ddot U \,=\,
V(\phi,e,m)e^{2U}\nn\,,\\
\frac{D^2 \phi^r}{D\tau^2} &\equiv & \ddot\phi^r +
\Gamma^r{}_{st}\dot\phi^s\,\dot\phi^t\,=\,
g^{rs}(\phi)\,\frac{\partial V(\phi,e,m)}{\partial \phi^s}
e^{2U}\,,\label{geoeq}
\eea
together with the constraint
\begin{equation}
\dot U^2+\frac{1}{2}\,g_{rs}(\phi)\,\frac{d\phi^r}{d\tau} \frac{d
\phi^s}{d\tau} - V(\phi,e,m)\,e^{2U}=c^2 \,,\label{bhconstr}
\ee
where the positive definite geodesic potential $V(\phi,e,m)$ has the
following form:
\begin{eqnarray}
V&=&-\frac{1}{2}\,\Gamma^T\mathcal{M}(\phi)\,\Gamma=-\frac{1}{2}\,\Gamma^T
\left(\matrix{I+R\,I^{-1}\,R & -R\,I^{-1}\cr -I^{-1}\,R &
I^{-1}}\right)\Gamma >0\,,\label{geopot}
\end{eqnarray}
in terms of the vector of magnetic-electric charges $\Gamma\equiv
(m^\Lambda,\,e_\Lambda)$. Here the dotted quantities are
differentiated with respect to the evolution parameter $\tau$.
\par

The equations of motion (\ref{geoeq}) can be associated to a one
dimensional theory whose Lagrangian\footnote{Note that here and in
the following by abuse of language we adopt the terms  Lagrangian
and Hamiltonian even if the evolution variable $\tau$ does not
describe the temporal evolution but the radial one.}
 is:
\begin{eqnarray}
\mathcal{L}=\dot{U}^2+\frac{1}{2}\,G_{rs}(\phi)\,\dot{
\phi}^r\,\dot{
\phi}^s+e^{2\,U}\,V(\phi)=\frac{1}{2}\,G_{ij}(q)\,\dot{q}^i\,\dot{q}^j+\mathcal{V}(q)\,,\label{lag}
\end{eqnarray}
where $\mathcal{V}(q)\equiv e^{2\,U}\,V(\phi)$ and we introduced the
functions $q^i(\tau)=(U(\tau),\,\phi^r(\tau))$ together with the
metric
\begin{eqnarray}
G_{ij}(q)&=&\left(\matrix{2 & {\bf 0}\cr {\bf 0} & G_{rs}(\phi)
}\right)\,.
\end{eqnarray}
Once the Lagrangian is known we can proceed with an Hamiltonian
approach using the phase space that stems from the
$q^i(\tau)=(U(\tau),\,\phi^r(\tau))$ variables, introducing the
conjugate momenta to $q^i$:
\begin{eqnarray}
p_i&=&\frac{\delta \mathcal{L}}{\delta
\dot{q}^i}=G_{ij}\,\dot{q}^j\,.\label{pq}
\end{eqnarray}
In  terms of the phase-space variables
 $q^i$ and $p_i$  the Hamiltonian $H(p,q)$ then reads:
\begin{eqnarray}
H(p,\,q)&=&\frac{1}{2}\,p_i\,G^{ij}\,p_j-\mathcal{V}(q)=
\frac{1}{2}\,\dot{q}^i\,G_{ij}(q)\,\dot{q}^j-\mathcal{V}(q)\,.
\end{eqnarray}
This is consistent with the constraint (\ref{bhconstr}) that
acquires the meaning of  ``energy conservation'':
\begin{eqnarray}
H(p,\,q)&=&c^2\,\,\Leftrightarrow\,\,\,\frac{1}{2}\,\dot{q}^i\,G_{ij}(q)\,\dot{q}^j-\mathcal{V}(q)=c^2\,.
\label{hconstr}\end{eqnarray}

\section{Hamilton--Jacobi Formalism and Static Black Holes}
\label{sec2} Let us recall how the solutions of the equations of
motion can be obtained  by applying the machinery of the
Hamilton--Jacobi theory. We consider the principal Hamiltonian
function $S(q,P,\tau)$ depending on $q^i$ and on new constant
momenta $P_i$. It is defined by the set of first order equations:
\begin{eqnarray}
\frac{\partial S}{\partial q^i}&=&p_i\,\,\,,\,\,\,\,\frac{\partial
S}{\partial P_i}=Q^i\,\,\,,\,\,\,\,\frac{\partial S}{\partial
\tau}=-H\,,\label{pf}
\end{eqnarray}
where $P_i,\,Q^i$ are new constant canonical variables which can be
expressed in terms of the initial values of  $q^i$ and $p_i$. From
the general theory of canonical transformations, see for instance
\cite{mh}, it is known that the above transformation generated by
$S$ always exists locally in the $p,\,q$ space, in a neighborhood of
any point which is not critical, namely in  which $(\frac{\partial
H}{\partial q},\, \frac{\partial H}{\partial p})\neq ({\bf 0},\,{\bf
0})$.\par
 We shall leave the dependence
of $S$ on the
constant $P_i$ implicit and focus on its dependence on the $q^i$.\\
From the last and the first equations of (\ref{pf}) and from the
Hamiltonian constraint \eq{hconstr} we have:
\begin{eqnarray}
\label{HJ1} S(q,\tau)&=&{\mathcal{W}(q)}-c^2\,\tau\,\\
p_i&=&\frac{\partial \mathcal W}{\partial q^i}\,,\label{spq}
\\
\label{HJ2}H(q,\partial_q \mathcal W) &\equiv &\frac{1}{2}\,\partial_i\mathcal W\,G^{ij}\,\partial_j
\mathcal W-\mathcal{V}(q)= c^2\,,
\end{eqnarray}
where \eq{HJ2} defines the Hamilton--Jacobi equation for the
function $\mathcal W$, usually called Hamilton characteristic
function.\par From eq. (\ref{spq}) and eq. (\ref{pq}) we finally get
\begin{eqnarray}
\label{dyn2}\dot{q}^i&=&G^{ij}\,p_j=G^{ij}\,\frac{\partial \mathcal
W}{\partial q^j}\,.
\end{eqnarray}
This shows that, provided a solution to the equations
(\ref{HJ1})-(\ref{HJ2}) is found, the evolution of the metric and
scalar fields can be described in terms of a dynamical system of the
form (\ref{dyn}). In the next section we give its unique
solution.\par Note that the function $S$ in eq. \eq{HJ1} generalizes
the expression for the prepotential conjectured in
\cite{Andrianopoli:2007gt} for the general (non-extremal) case.
To make contact with the proposal in \cite{Andrianopoli:2007gt}, let
us consider the following expression for the principal function $S$:
\begin{eqnarray}
S(U,\phi^r,\tau)=2\,e^U
W(\phi^r,\tau)+c^2\,\tau=\mathcal{W}(U,\,\phi^r)-c^2\,\tau\,.\label{rame}
\end{eqnarray}
This expression reproduces the first order equations for the
prepotential as given in
\cite{Ceresole:2007wx,Andrianopoli:2007gt}:
\begin{equation}\label{fo}
{\dot{U}=e^U\, {W}\,,\qquad\dot{\phi}^r=2\,e^U\,G^{rs}\,\partial_s
{W} }\,,\label{1oreqs}
\end{equation}
together with the condition found in \cite{Andrianopoli:2007gt}
for the non-extremal case: $
\partial_\tau W = -c^2 e^{-U}$.\par
We observe, however, that \eq{rame} is a rather restrictive Ansatz
for the Hamilton's principal function, since the dependence of
$\mathcal{W}$ on $U$ may not have a factorized form, as it is
apparent, for example, in the Reissner--Nordstrom case discussed
below. Actually, the present discussion generalizes such an Ansatz.
\subsection{Critical points as attractors.} The first order system
(\ref{dyn2}) may have critical points in the space of the $q^is$,
namely points $(q_0^i)$ in which its right hand side vanishes:
$\partial_i \mathcal{W}_{\vert q_0}=0$. From the Hamilton--Jacobi
equation it is clear that such points may exist only in the extremal
case, since the right hand side of the equation
\begin{eqnarray}
\partial_i\mathcal{W}\,G^{ij}\,\partial_j \mathcal{W}=2\,(c^2+e^{2\,U} \,V(\phi))\,,
\end{eqnarray}
being $V\ge 0$, may vanish only if $c=0$. A critical point is an
\emph{attractor} \cite{attractor} if at that point the Hessian of the geodesic
potential is positive. Attractors are reached at
$\tau\rightarrow -\infty$,  where $U(-\infty)\to -\infty$, which
corresponds to the horizon, since, for $c=0$,
$\tau=-1/(r-r_0)$.\par In the extremal case we can make the Ansatz
$\mathcal{W}(U,\,\phi^r)=2\,e^U\, W(\phi^r)$, which corresponds to
taking $W$ in eq. (\ref{rame}) not to depend explicitly on
$\tau$, and the Hamilton--Jacobi equation translates in the
following equation for $W(\phi^r)$:
\begin{eqnarray}
W^2+2\,G^{rs}\,\frac{\partial W}{\partial \phi^r}\,\frac{\partial
W}{\partial \phi^s}&=&V(\phi^r)\,.\label{hjw0}
\end{eqnarray}
The ADM mass of the solution is given by the value of $\dot{U}$ at
radial infinity ($r\rightarrow \infty,\,\tau\rightarrow 0^-$):
$M_{ADM}(\phi^r_\infty,\,e,m)=lim_{\tau\rightarrow 0^-}\,\dot{U}$, where
$\phi_\infty^r$ are the boundary values of the scalar fields at infinity.
From the first order equations (\ref{1oreqs}) written in terms of
$W(\phi^r,\,e,m)$, we see that, since $\lim_{\tau\rightarrow
0^-}\,e^U=1$,  the ADM mass of the solution coincides with the value
of $W$ as a function of the values of $\phi^r_\infty$:
$M_{ADM}(\phi^r_\infty,\,e,m)=W(\phi^r_\infty,\,e,m)$.
\paragraph{Multiple attractors} For a given set of quantized electric and magnetic charges
$\Gamma=(m^\Lambda,\,e_\Lambda)$, extremal solutions may have more
than one attractors \cite{moore,basins}. In \cite{moore} a necessary
condition was given for the existence of multiple attractors. We want
here to give a general proof of this statement for generic (not
necessarily BPS) attractors. As in \cite{attractor2} we can define
the ``Weinhold metric'' as the Hessian of
$M_{ADM}(\phi^r_\infty,\,e,m)=W(\phi^r_\infty,\,e,m)$ with respect to the
boundary values $\phi^r_\infty$:
\begin{eqnarray}
\hat{\bf W}& =&(W_{rs})\equiv \left(\frac{\partial^2
W}{\partial\phi^r\,\partial\phi^s}\right)_{\phi=\phi_\infty}\,.
\end{eqnarray}
We can take $\phi_\infty^r$ to coincide with the critical point $\phi_0^r$. In this
case $\partial_r V (\phi_0^r)=\partial_r W (\phi_0^r)=0$ and the
scalar fields of the solution will not evolve: $\phi^r(\tau)\equiv
\phi_0^r$. This solution is called \emph{double extremal}. By
computing the second derivatives of eq. (\ref{hjw0}) we can relate the Hessian matrix $
(W_{rs})$ of $W$ to the Hessian matrix  $\hat{\bf V}\equiv
(\partial_r\partial_s\,V)$ of $V$ at a critical point as follows:
\begin{eqnarray}
2\,\sqrt{V}\,\hat{\bf W}+4\,\hat{\bf W}\,\hat G^{-1}\,\hat{\bf W}&=&\hat {\bf V}\,,
\end{eqnarray}
where $\hat G^{-1}\equiv (G^{rs})$. The above matrix equation is solved by
\begin{eqnarray}
\hat{\bf W}\,\hat G^{-1}&=&\frac{\sqrt{V}}{4}\left(-\bfone+\sqrt{\bfone+\frac{4}{V}\,\hat{\bf
V}\,\hat G^{-1}}\right)\,,\label{WV}
\end{eqnarray}
 where $\bfone \equiv (\delta_r^s)$. The matrix $\hat{\bf
V}\,\hat G^{-1}=(\partial_r\partial_s V\,G^{s\ell})_{\vert \phi_0}$ is
the ``mass squared'' matrix of the fluctuations of $\phi^r$ around
$\phi_0^r$. We can consider for instance BPS black holes in
$\mathcal{N}=2$ supergravity, described by a function $W=|Z|$, $Z$
being the complex central charge of the theory. At the corresponding
extremum one can show that $\partial_r\partial_s V(\phi_0^r)=2\,V\,
G_{rs}(\phi_0^r)>0$ \cite{attractor2}, and from eq. (\ref{WV}) one
finds
\begin{eqnarray}
\partial_r\partial_s W(\phi_0^r)&=&\frac{1}{4\,\sqrt{V}}\,\partial_r\partial_s
V(\phi_0^r)=\frac{\sqrt{V}}{2}\,G_{rs}(\phi_0^r)>0\,\,\,\,\,\,\,\mbox{(BPS
attractor)}\,\,\,.
\end{eqnarray}
In general a critical point $\phi_0^r$ is an attractor iff $\hat{\bf
V}(\phi_0)$, or equivalently $\hat{\bf V}\,\hat G^{-1}(\phi_0)$, are positive definite.
Eq. (\ref{WV}) then implies that $\phi_0^r$ is an attractor iff
$\hat{\bf W}\,\hat G^{-1}(\phi_0)$, or equivalently $\hat{\bf W}(\phi_0)$, are positive definite. 
If, for a given choice of the quantized charges $e,\,m$, $W$
(or equivalently $V$) has more than one attractor point, $W$ must
have more than one minimum and this implies that there should exist
at least one critical point in which the Hessian of $W$ has negative
eigenvalues. Therefore \emph{if the Weinhold metric is positive
definite everywhere in the moduli space, there can be no more than
one attractor} which is the statement made in \cite{moore}.
\section{Solution to the Hamilton--Jacobi equation}\label{sec25}
We recall that the solution of the set of differential equations
(\ref{pf}), \eq{HJ2} in terms of Hamilton's principal function $S$ is formally
given by (see e.g. \cite{arnold})
\begin{eqnarray}
S(q,\tau)&=& S_0+
 \int_{q_0,\tau_0}^{q,\tau}\mathcal{L}(q,\,\dot{q})\,d\tau\,,\label{ssol}
\end{eqnarray}
where the integral is performed along the characteristic trajectory
$\gamma=q^i(\tau)$, i.e. the solution of  Hamilton's equations, such that:
\begin{eqnarray}
q^i(\tau_0)&=&q^i_0\,\,\,\,,\,\,\,\,\,q^i(\tau)=q^i\,.\label{interpolate}
\end{eqnarray}
 Few comments
here are in order. The above formula provides, in the most general
case, only a \emph{local} definition of $S$: local in $\tau$, to
avoid multivaluedness of $S$ \cite{arnold}, and local in the configuration space
with coordinates $q^is$, being $S$ defined only on the points $(q^i)$,
for fixed $q_0^i,\,\tau_0$, for which the interpolating
characteristic trajectory, satisfying  (\ref{interpolate}), exists.
In fact, as pointed out in Section \ref{sec2}, locally, in the
neighborhood of a non-critical point in the phase space, there
always exist a \emph{complete} solution $S(q^i,\,P_i,\,\tau)$ to the
Hamilton--Jacobi equation and it has the form (\ref{ssol}) ($P_i$
can be seen as a complete set of integration constants).
In what follows, we shall use eq. (\ref{ssol}) bearing its local
validity in mind.
\par In our problem, we can use the Hamiltonian constraint to find
the expression of the principal function in terms of the potential,
as follows:
\begin{eqnarray}
S(q,\tau)&=&S_0+
\int_{q_0,\tau_0}^{q,\tau}\left[c^2+2\,\mathcal{V}(q)\right]\,d\tau\,,
\end{eqnarray}
so that, using \eq{lag} and \eq{hconstr}, the Hamilton's characteristic function is given by:
\begin{eqnarray}
\mathcal{W}(q)&=&\mathcal{W}_0+
\int_{q_0,\tau_0}^{q,\tau}\left[c^2+\mathcal{L}(q,\,\dot{q})\right]\,d\tau=\mathcal{W}_0+2\,\int_{q_0,\tau_0}^{q,\tau}\left[c^2+\mathcal{V}(q)\right]\,d\tau\,.\label{Wfun}
\end{eqnarray}
Actually, the above formula can also be derived from direct integration of eq.
(\ref{HJ2}). Indeed  (\ref{HJ2}) has the form of the eikonal equation
for a wave front $\mathcal W= const.$ propagating in a medium of
refractive index $n=\sqrt{2(c^2+\mathcal V)}$:
\begin{equation}\label{eik}
   n^2=\partial_i\mathcal W\,G^{ij}\,\partial_j
\mathcal W.
\end{equation}
From equation (\ref{dyn2}) we get that $\partial_i \mathcal{W} $ is
tangent to the ``light rays" namely the characteristics
$\gamma=(q_i(\tau))$. Introducing the  proper distance $s$ along a
characteristic:
\begin{equation}\label{ds}
  ds=\sqrt{\dot{q}^i\,G_{ij}(q)\,\dot{q}^j}\,d\tau =\sqrt{2(c^2+\mathcal{V}(q))}\,d\tau
\end{equation}
using \eq{dyn2} we have:
\begin{equation}\label{ds1}
\frac{d \mathcal W}{ds}=\frac{\partial \mathcal W
}{dq_i}\frac{dq_i}{d\tau}\frac{d\tau}{ds}
=\partial_i\mathcal{W}\,G^{ij}\,\partial_j \mathcal{W}\,\frac{d\tau}{ds}
=\sqrt{2(c^2+\mathcal{V}(q))}
\end{equation}
that is:
\begin{equation}\label{dw}
    d\mathcal{W}=\sqrt{2(c^2+\mathcal{V}(q))}\,ds=2(c^2+\mathcal{V}(q))\,d\tau.
\end{equation}
From the above equation it follows that $\frac{d\mathcal{W}}{d\tau}$
along $\gamma$ is positive, namely that $\mathcal{W}$ is a monotonic
increasing function of $\tau$ along a solution (the same is true for
the principal function $S$, since the Lagrangian is non-negative).
\paragraph{Example} Let us review the construction of $\mathcal{W}$ for the
Reissner-Nordstr\"om black hole \cite{Janssen:2007rc}. The $q^i$
variables now consist of the function $U$ alone. This is for
instance a solution to $\mathcal{N}=2$ pure supergravity. With
respect to the only vector field of the theory (the graviphoton) the
solution can have in general an electric and a magnetic charge
$e,\,m$. The geodesic potential reads:
\begin{eqnarray}
\mathcal{V}(U,\,e,\,m)&=&e^{2\,U}\,Q^2\,\,\,\,,\,\,\,\,\,Q^2\equiv
\frac{1}{2}(e^2+m^2)\,.
\end{eqnarray}
The Hamiltonian constraint and the Hamilton--Jacobi equation read:
\begin{eqnarray}
\dot{U}^2&=&(\partial_U \mathcal{W})^2=c^2+e^{2\,U}\,Q^2\,.
\end{eqnarray}
We can then readily apply eq. (\ref{Wfun}) to find, upon changing
variables from $\tau$ to $U$:
\begin{eqnarray}
\mathcal{W}(U)&=&
\mathcal{W}_0+2\,\int_{U_0,\tau_0}^{U,\tau}\left[c^2+e^{2\,U}\,Q^2\right]\,d\tau=
\mathcal{W}_0+2\,\int_{U_0}^{U}\left[c^2+e^{2\,U}\,Q^2\right]\,\frac{dU}{\dot{U}}=\nonumber\\&=&
\mathcal{W}_0+2\,\int_{U_0}^{U}\sqrt{c^2+e^{2\,U}\,Q^2}\,dU=\nonumber\\&=&
\mathcal{W}_0+2\,\left[\sqrt{c^2+e^{2\,U}\,Q^2}-\frac{c}{2}\,\,
\log\left(\frac{\sqrt{c^2+e^{2\,U}\,Q^2}+c}{\sqrt{c^2+e^{2\,U}\,Q^2}-c}\right)\right]\,.
\end{eqnarray}
\section{Duality invariance of the prepotential
$\mathcal{W}$.}\label{sec3} Let us now consider an extended
supergravity theory in $D=4$. It is known that the global symmetries
of the equations of motion and Bianchi identities are encoded in the
isometry group $G$ of the scalar manifold (if non-empty), whose
action on the scalar fields is associated with a simultaneous linear
symplectic action on the field strengths $F^\Lambda$ and their duals
$G_\Lambda$. The duality action of $G$ is defined by an embedding
$D$ of $G$ inside the group ${\rm Sp}(2\,n_v,\mathbb{R})$:
\begin{eqnarray}
g\in G&:&\cases{\phi^r\rightarrow \phi^{r\,\prime}=g\star
\phi^r\cr \left(\matrix{F^\Lambda\cr G_\Lambda}\right)\rightarrow
D(g)\cdot \left(\matrix{F^\Lambda\cr G_\Lambda}\right) }\,,
\end{eqnarray}
where $g\star$ denotes the non-linear action of $g$ on the scalar
fields and $D(g)$ is the $2\,n_v\times 2\,n_v$ symplectic matrix
associated with $g$.

We are going to prove explicitly that the prepotential
$\mathcal{W}(q)$ is invariant under the duality action of $G$.
Keeping in mind that the  metric (and therefore the function $U$) is
a duality invariant field, we define $(g\star q^i)\equiv (U,\,g\star
\phi^r)$. The on-shell global  invariance of the four dimensional
theory under $G$ implies that, if $\gamma=(q^i(\tau))$ is a
characteristic trajectory of the Lagrangian system (\ref{lag}) with
charge parameters $\Gamma=(m^\Lambda,\,e_\Lambda)$, then
$g\star\gamma=(g\star q^i(\tau))$ is a trajectory of the Lagrangian
system (\ref{lag}) with charge parameters $D(g)\cdot\Gamma$.
\par Let us make the dependence of the geodesic potential
$\mathcal{V}$ on the electric-magnetic charges explicit by writing
$\mathcal{V}(q,\,\Gamma)$. From general properties of the
symplectic matrix $\mathcal{M}(\phi)$, defined in (\ref{geopot}),
we have:
\begin{eqnarray}
\mathcal{M}(g\star
\phi)&=&D(g)^{-T}\,\mathcal{M}(\phi)\,D(g)^{-1}\,.
\end{eqnarray}
From this it follows that the potential $\mathcal{V}$ is duality
invariant, in the sense that:
\begin{eqnarray}
\mathcal{V}(g\star q,\,D(g)\cdot \Gamma)&=&\mathcal{V}(
q,\,\Gamma)\,.\label{Vinv}
\end{eqnarray}
The group $G$ is then a global symmetry group of the one-dimensional
Lagrangian $\mathcal{L}$ in (\ref{lag}) and thus of both Hamilton's
principal and characteristic functions $S(q,\,\tau;\,\Gamma)$ and
$\mathcal{W}(q;\,\Gamma)$. Indeed from (\ref{Wfun}) and (\ref{Vinv})
we have:
\begin{eqnarray}
\mathcal{W}(q;\,\Gamma)&=&\mathcal{W}_0+2\int_{q_0,\,\tau_0}^{q,\,\tau}\left[c^2+\mathcal{V}(q,\,\Gamma)\right]\,d\tau=
\mathcal{W}_0+2\int_{g\star q_0,\,\tau_0}^{g\star
q,\,\tau}\left[c^2+\right.\nonumber\\&&+\left.\mathcal{V}(g\star
q,\,D(g)\cdot \Gamma)\right]\,d\tau=\mathcal{W}(g\star q;\,D(g)\cdot
\Gamma)\,.
\end{eqnarray}

\section{The extended phase space and time-reductions to
$D=3$}\label{sec4} Four-dimensional static and spherically symmetric
black holes depend on a number of degrees of freedom which includes,
besides the metric and scalars, also the degrees of freedom
corresponding to the gauge potentials. The one-dimensional effective
Lagrangian \eq{lag} encoding the information on the theory can then
been formulated as a Lagrangian for a geodesic model describing all
the degrees of freedom, as discussed in
\cite{breit},\cite{attractor2}. It is then natural to extend the
phase space to include as further degrees of freedom the magnetic
and electric potentials together with their conjugate momenta. This
approach was pionereed in \cite{canonically}, in the case of double
extremal black holes.\par As it is well known, this approach is
equivalent, for static black holes, to a time reduction of the four
dimensional Lagrangian (\ref{S4}) \cite{breit}.\par
 To implement the reduction on the time direction, let us consider  the following Ansatz for the metric
\cite{breit} \footnote{This metric describes stationary solutions,
in the static case $A^0_i=0$.}:
\begin{eqnarray}
ds_4^2&=&e^{2U}\,(dt+A^0_i\,dx^i)^2-e^{-2U}\,g^{(3)}_{ij}\,dx^i\,dx^j\,,
\end{eqnarray}
where $x^i$, $i=1,2,3$, are the coordinates of the final Euclidean
space. In this theory all the vectors are dualized to scalar fields so
as to obtain a sigma model coupled to gravity. Let us introduce the
$n_v$ three dimensional scalars $\zeta^\Lambda=A_0^\Lambda$ which,
together with the scalars $\tilde{\zeta}_\Lambda$, dual in $D=3$ to
the vectors $A^\Lambda_i$, form the symplectic vector of electric
and magnetic potentials
$\Z^M=(\zeta^\Lambda,\,\tilde{\zeta}_\Lambda)$. Finally we shall
denote by $a$ the axion dual in $D=3$ to the Kaluza-Klein vector
$A^0_i$. The final $D=3$ action reads:
\begin{eqnarray}
S_3&=&\int
d^3x\,\sqrt{|g^{(3)}|}\,\left(\frac{1}{2}\,R_3-\frac{1}{2}\,G_{IJ}(\Phi)\,\partial_i\Phi^I\,\partial^i\Phi^J\right)\,,\nonumber
\end{eqnarray}
where $\Phi^I\equiv(U,\phi^r,\,a,\,\Z^M)$, and  the sigma model
metric reads:
\begin{eqnarray}
G_{IJ}(\Phi)\,d\Phi^I\,d\Phi^J&=&
dU^2+\frac{1}{2}\,G_{rs}\,d\phi^r\,d\phi^s+\frac{e^{-4\,U}}{4}\,
\omega^2+\frac{e^{-2\,U}}{2}\,d\Z^T\,\mathcal{M}\,d\Z\,,\nonumber
\end{eqnarray}
where $\mathcal{M}_{MN}$ is the negative definite matrix defined
 in (\ref{geopot}) and the one-form $\omega$ is defined as $\omega=da-\Z^T\,\mathbb{C}\,d\Z$, $\mathbb{C}$ being  the
 antisymmetric $2\,n_v\times 2\,n_v$ ${\rm Sp}(2\,n_v,\,\mathbb{R})$-invariant
 matrix, for which we shall use  the following
form
\begin{equation}
\mathbb{C} =\pmatrix{0&-\bfone \cr \bfone&0}
\,.
\end{equation}
The scalar fields $\Phi^I$ span a pseudo-Riemannian manifold ${\Scr
 M}_3$ which contains as a submanifold the scalar manifold ${\Scr
 M}_4$ of the $D=4$ parent theory spanned by $\phi^r$.\par
 Spherically symmetric black-hole solutions are described by
 \emph{geodesics} $\Phi^I(\tau)$ on ${\Scr M}_3$ para\-me\-trized by the radial variable $\tau$ \cite{breit}.
 We shall restrict ourselves to spherically symmetric
 solutions with vanishing NUT charge, namely we shall take $\omega_\tau=0$. These solutions will be
 described by the following effective Lagrangian:
 \begin{eqnarray}
\mathcal{L}_3&=&\dot{U}^2+\frac{1}{2}\,G_{rs}\,\dot{\phi}^r\,\dot{\phi}^s+\frac{e^{-2\,U}}{2}\,\dot{\Z}^T\,
\mathcal{M}\,\dot{\Z}=\frac{1}{2}\,G_{\alpha\beta}\,\dot{q}^\alpha\,\dot{q}^\beta\,.\nonumber
 \end{eqnarray}
 Let us introduce the following generalized coordinates
 $q^\alpha\equiv(U,\phi^r,\,\Z^M)=(q^i,\Z^M)$. The
 conjugate momenta $p_\alpha$ will read:
 \begin{eqnarray}
p_\alpha=G_{\alpha\beta}(q^i)\,\dot{q}^\alpha\,=(p_i,p_M)\,.
 \end{eqnarray}
 In terms of $q^\alpha,\,p_\alpha$ we write the Hamiltonian:
 \begin{eqnarray}
H_3(p,q)&=&\frac{1}{2}\,G^{\alpha\beta}\,p_\alpha\,p_\beta=\frac{1}{2}\,G^{ij}(q^i)\,p_i\,p_j+\frac{e^{2\,U}}{2}\,p_M\,
\mathcal{M}^{MN}(\phi^r)\,p_N=c^2\,,\nonumber
 \end{eqnarray}
 where $\mathcal{M}^{MN}$ denotes the inverse matrix of
 $\mathcal{M}_{MN}$, given by:
 $\mathcal{M}^{MN}=\mathbb{C}^{MP}\,\mathbb{C}^{NQ}\,\mathcal{M}_{PQ}$.
 Since $\Z^M$ are cyclic, the corresponding momenta $p_M$ are constants of
 motion. They are identified with the quantized charges as
 follows: $p_M=-\mathbb{C}_{MN}\,\Gamma^N$. With this
 identification, the last term in $H_3$ reads $\frac{1}{2}\,e^{2\,U}\,\Gamma^T\cdot \mathcal{M}\cdot \Gamma=-\mathcal{V}(q^i,\Gamma)$
 and $H_3$ coincides with the Hamiltonian $H$ defined in the
 previous sections. Therefore the resulting  equations of motion for $p_i,\,q^i$ are the same as those discussed earlier.
   As far as the equation for $\Z^M$ is concerned, it reads \cite{breit}:
   \begin{eqnarray}
\dot{\Z}^M&=&\frac{\partial H_3}{\partial
p_M}=-e^{2\,U}\,\mathbb{C}^{MN}\,\mathcal{M}_{NP}\,\Gamma^P\,.\label{esposito}
   \end{eqnarray}

    This analysis can be viewed
   as an extension of that given in \cite{canonically} since it includes
in the definition of the phase space also the four dimensional
scalar fields.\par In this enlarged Hamiltonian system we want now
to define Hamilton's characteristic function, to be denoted by
$\mathcal{W}_3(q^\alpha,\,P_\alpha)$. By definition this function
generates the canonical transformation to the coordinates
$Q^\alpha,\,P_\alpha$, where $P_\alpha$ are constants of motion.
Since also $c^2$ is conserved, it will be a function of the
$P_\alpha$, $c^2=c^2(P)$. It will indeed provide the Hamiltonian
in the new coordinates. From the general theory it is known that
the coordinates $Q^\alpha$ are linear in $\tau$, i.e. harmonic
functions:
\begin{eqnarray}
Q^\alpha &=& \left(\frac{\partial c^2}{\partial
P_\alpha}\right)\,\tau+Q_0^\alpha\,.
\end{eqnarray}
If we choose one of the $P_i$ to coincide with $c^2$, then only
the corresponding $Q^i$ will be linear in $\tau$, the other
$Q^\alpha$ being constants. The function
$\mathcal{W}_3(q^\alpha,\,P_\alpha)$ satisfies the following
relations:
\begin{eqnarray}
p_\alpha&=&\frac{\partial \mathcal{W}_3}{\partial q^\alpha}\,,\label{eq11}\\
Q^\alpha&=&\frac{\partial \mathcal{W}_3}{\partial P_\alpha}\,,\label{eq12}\\
c^2&=&H_3(q^\alpha,\,\frac{\partial \mathcal{W}_3}{\partial
q^\alpha})\,,\label{eq13}
\end{eqnarray}
the latter being the Hamilton--Jacobi equation. Since $p_M$ are
already constant, $\mathcal{W}_3$ should be such that $P_M=p_M$.
This function can be expressed in terms of the four-dimensional
Hamilton's characteristic function $\mathcal{W}$ as follows:
\begin{eqnarray}
\mathcal{W}_3(q^\alpha,\,P_\alpha)&=&
\mathcal{W}(q^i,\,P_i,\,P_M)+\Z^M\,P_M\,,\label{w3w}
\end{eqnarray}
where $\mathcal{W}$ was defined in (\ref{HJ2}). Equation
(\ref{eq11}), for $\alpha=i$, follows from (\ref{spq}) and, for
$\alpha=M$ implies $p_M=P_M$. Therefore the dependence of
$\mathcal{W}$ on $P_M$ is nothing but the dependence of the
four-dimensional $\mathcal{W}$ on the quantized charges
\begin{equation}
\Gamma^M=\mathbb{C}^{MN}\,p_N=\mathbb{C}^{MN}\,P_N\,.\label{gammap}
\end{equation}
 Equation
(\ref{w3w}) can then be rewritten in the form:
\begin{eqnarray}
\mathcal{W}_3(q^\alpha,\,P_\alpha)&=&
\mathcal{W}(q^i,\,P_i,\,\Gamma^M)-\Z^M\,\mathbb{C}_{MN}\,\Gamma^N\,.
\end{eqnarray}
 Let us now
consider the component $\alpha=M$ of eq. (\ref{eq12}):
\begin{eqnarray}
\frac{\partial \mathcal{W}_3}{\partial P_M}&=&\frac{\partial
\mathcal{W}}{\partial P_M}+\Z^M=Q^M\,.
\end{eqnarray}
The above equation can also be written, using \eq{gammap}:
\begin{eqnarray}
\Z^M-\mathbb{C}^{MN}\,\frac{\partial \mathcal{W}}{\partial
\Gamma^N}&=&Q^M\,.
\end{eqnarray}
This is a non-trivial equation which implies that the combination
on the left hand side is a symplectic vector of harmonic
functions. Since the  $Q^M$ can be chosen to be constant,  we can
write:
\begin{eqnarray}
\Z^M&=&\mathbb{C}^{MN}\,\frac{\partial \mathcal{W}}{\partial
\Gamma^N}+const\,.\label{eqz}
\end{eqnarray}
The above equation allows to compute the electric-magnetic
potentials once the $\mathcal{W}$- prepotential is known on the
solution as a function of all the quantized charges.\par We shall
check  below eq. (\ref{eqz}) on some specific solutions.
\paragraph{The BPS Solution for the $\mathcal{N}=2$ Case.\\}
We shall refer to the usual $\mathcal{N}=2$ special geometry
notations. Let $z^i$ denote the complex scalar fields on the special
K\"ahler manifold and let $V^M(z,\bar{z}),\,U_i^M(z,\bar{z})$ be the
covariantly holomorphic symplectic section and its covariant
derivative:
\begin{eqnarray}
V^M &=&\pmatrix{L^\Lambda\cr M_\Lambda}\,\,\,,\,\,\,\,\,U^M_i =D_i
V^M=\pmatrix{f_i^\Lambda\cr h_{\Lambda i}}\,.\label{vu}
\end{eqnarray}
 The matrix $\mathcal{M}=(\mathcal{M}_{MN})$ is related to the above quantities as
follows:
\begin{eqnarray}
\mathbb{C}\,\mathcal{M}\,\mathbb{C}&=&-\mathcal{M}^{-1}=V\,\bar{V}^T+\bar{V}\,V^T+g^{i\jbar}\,U_i\,\bar{U}_{\jbar}^T+
g^{\jbar i}\,\bar{U}_{\jbar}\,U_i^T\,.\label{mvu}
\end{eqnarray}
 The symplectic section $V^M$ also satisfies the property:
$V^T\,\mathbb{C}\,V=-{\rm i}$.\par
 The central charge $Z$ is defined as follows:
\begin{eqnarray}
Z&=&\Gamma^T\,\mathbb{C}\,V=e_\Lambda\,L^\Lambda-m^\Lambda\,M_\Lambda\,,
\end{eqnarray}
 The first order
equations describing the spatial evolution of the BPS solution
originate from the Killing-spinor equations and read:
\begin{eqnarray}
\dot U &=& e^U\, |Z|\,\,,\,\,\, \dot z^i =2 e^U\,g^{i\jbar}\,
\partial_\jbar |Z|\,. \label{ks}
\end{eqnarray}
The corresponding prepotential $\mathcal{W}$ has the following form
$ \mathcal{W}=2\,e^U\,|Z|.$ \par We wish now to verify equation
(\ref{eqz}) for this class of solutions. To this end we show that
the derivative of the right hand side of this equation equals
$\dot{\mathcal{Z}}$, as given from eq. (\ref{esposito}):
\begin{eqnarray}
\frac{d}{d\tau}\frac{\partial
\mathcal{W}}{\partial \Gamma^M}&=&-\mathbb{C}_{MN}\,\dot{\mathcal{Z}}^N=-e^{2\,U}\,\mathcal{M}_{MN}\,\Gamma^N\,.\label{ZdotW}
\end{eqnarray}
Let us define the quantity:
\begin{eqnarray}
T&=&H^T\,\mathbb{C}\,V=H_\Lambda\,L^\Lambda-H^\Lambda\,M_\Lambda\,,
\end{eqnarray}
where we have introduced the symplectic vector $H^M$ of harmonic
functions $ H^M(\tau)\equiv h^M-\sqrt{2}\,\Gamma^M\,\tau$. In
terms of the above quantities, it was shown in
\cite{Behrndt:1998eq} that the BPS solution is defined by the
 following algebraic equations:
 \begin{eqnarray}
\bar{T}\,V^M-T\,\bar{V}^M&=&-\frac{{\rm
i}}{\sqrt{2}}\,H^M\,\,\,,\,\,\,\,\,e^{-U}=|T|\,,\label{54}
 \end{eqnarray}
with the condition that $H^T\,\mathbb{C}\,\dot{H}=0$. From the above
relations and positions one can prove the following properties:
\begin{eqnarray} {\rm Im}(T\,\bar{Z})=0\,\,\,,\,\,\,\,
\dot{T}=-Z\,.\label{props}\end{eqnarray} Differentiating $
\mathcal{W}=2\,e^U\,|Z|$  with respect to $\Gamma$ and using
(\ref{eqz}) we find
\begin{eqnarray}
\mathcal{Z}^M&=&-2\,\frac{e^{U}}{|Z|}\,{\rm Re}(\bar{Z}\,V^M)=-2\,e^{2\,U}\,{\rm Re}(\bar{T}\,V^M)\,.
\end{eqnarray}
Using (\ref{ks}) and (\ref{props}) one finds:
\begin{eqnarray}
\frac{d}{d\tau}\,(\bar{T}\,V^M)&=&\left(V^M\,\bar{V}^N-g^{i\,\jbar}\,U_i^M\,\bar{U}_{\jbar}^N\right)\,\mathbb{C}_{NP}\,\Gamma^P\,,
\end{eqnarray}
and then
\begin{eqnarray}
\dot{\mathcal{Z}}^M&=&-4\,|Z|\,e^{3\,U}\,{\rm
Re}(\bar{T}\,V^M)-e^{2\,U}\,(V^M\,\bar{V}^N+\bar{V}^M\,{V}^N-g^{i\,\jbar}\,U_i^M\,\bar{U}_{\jbar}^N-\nonumber\\&&-g^{\jbar\,i}\,
\bar{U}_{\jbar}^M\,{U}_{i}^N)\,\mathbb{C}_{NP}\,\Gamma^P\,.\label{zp}
\end{eqnarray}
Using \eq{54},
the first term on the right
hand side of the above formula can be rewritten as follows
\begin{eqnarray}
-4\,|Z|\,e^{3\,U}\,{\rm
Re}(\bar{T}\,V^M)&=&2\,e^{2\,U}\,(V^M\,\bar{V}^N+\bar{V}^M\,{V}^N)\,\mathbb{C}_{NP}\,\Gamma^P\,.
\end{eqnarray}
Finally, using (\ref{mvu}) and the above property, equation
(\ref{zp}) then yields equation (\ref{ZdotW}).
\paragraph{The non-BPS extremal solution in $\mathcal{N}=2$ supergravity with $|Z|\neq 0$.\\}
We shall comment on an interesting class of non-BPS extremal
solutions in $\mathcal{N}=2$ supergravity. These solutions are
characterized by a non vanishing value of the central charge $Z$ at
the origin. An explicit form of this solution was worked out in
\cite{Hotta:2007wz,Gimon:2007mh,marrani2} for the one-modulus cubic
model and the STU model. In those two models the solutions are
characterized by three and five independent parameters respectively,
which are the duality invariant quantities that can be built out of
the central and matter charges. In \cite{Andrianopoli:2007gt} a
simple, duality invariant, form for the $\mathcal{W}$ prepotential
was given in terms of all the charges, which, however, did not
describe the full duality orbit of solutions. In
\cite{Ceresole:2007wx,LopesCardoso:2007ky} a relatively simple form
for $\mathcal{W}$ is given for cubic models as a function of
specific sets of the quantized charges. The corresponding solutions
however seem to exhibit all the duality invariant parameters. Let us
consider, for the sake of simplicity, the one-modulus $z^3$-model,
describing the supergravity multiplet coupled to a vector multiplet.
The bosonic sector consists in one complex scalar field $z=x-{\rm
i}\,y$, ($y>0$) and two vectors $A^\Lambda_\mu=(A^0_\mu,\,A^1_\mu)$,
$n_v=2$. The components of the  covariantly holomorphic section
$V^M$ in (\ref{vu}) read
\begin{eqnarray}
L^\Lambda &=& e^{\frac{K}{2}}\,(1,z)\,\,\,\,,\,\,\,\,M_\Lambda =
e^{\frac{K}{2}}\,(-z^3,z^2)\,\,\,,\,\,\,\,e^{\frac{K}{2}}=\frac{1}{2\,\sqrt{2}\,y^{\frac{3}{2}}}\,.
\end{eqnarray}
The matrix $\mathcal{M}_{MN}$ has the following form:
\begin{eqnarray}
\mathcal{M}_{MN}&=&\left(\matrix{ -\frac{{\left( x^2 + y^2 \right)
}^3}{y^3}   & \frac{3\,x\,
     {\left( x^2 + y^2 \right) }^2}{y^3} & - \frac{x^3}{y^3}  & -\frac{x^2\,
       \left( x^2 + y^2 \right) }{y^3}  \cr \frac{3\,x\,{\left( x^2 + y^2 \right) }^2}
   {y^3} & \frac{-3\,\left( 3\,x^4 + 4\,x^2\,y^2 + y^4 \right) }{y^3} & \frac{3\,x^2}{y^3} &
    \frac{3\,x^3 + 2\,x\,y^2}{y^3} \cr - \frac{x^3}{y^3}  & \frac{3\,x^2}
   {y^3} & -y^{-3} & -\frac{x}{y^3}   \cr -\frac{x^2\,
       \left( x^2 + y^2 \right) }{y^3} & \frac{3\,x^3 + 2\,x\,y^2}{y^3} & -
      \frac{x}{y^3}  & \frac{-\left( 3\,x^2 + y^2 \right) }{3\,y^3} \cr
      }\right)\,.\label{mmdd}
\end{eqnarray}
In \cite{Ceresole:2007wx} the prepotential $W$, related to
$\mathcal{W}$ by $W(\phi^r)=e^{-U}\,\mathcal{W}/2$, was given by:
\begin{eqnarray}
W_0&=&W_{\vert e_1,\,m^0=0}=e^{\frac{K}{2}}\,\vert
e_0-3\,m^1\,|z|^2\vert\,,\label{w01}\\
W_0&=&W_{\vert e_0,\,m^1=0}=e^{\frac{K}{2}}\,\vert
z\,(e_1+\,m^0\,|z|^2)\vert\,,\label{w02}
\end{eqnarray}
for the cases $(e_1,\,m^0)=0$ or  $(e_0,\,m^1)=0$, respectively.

Using $W_0$ we cannot fully check eq. (\ref{eqz}) since we do not
know the derivatives of $W$ along the charges which are set
to zero. We can however check those components of eq.
(\ref{eqz}) involving the derivatives of $W$ on the remaining
charges. Let us consider the $(e_1,\,m^0)=0$ case. The solution
exists for $e_0\,m^1<0$. Let us take $e_0<0,\,m^1>0$.
 The first order equations (\ref{1oreqs}), defined from $W_0$, read
 \begin{eqnarray}
\dot{U}&=&-\frac{e^U}{2\,\sqrt{2}\,y^{\frac{3}{2}}}\,\left(e_0-3\,m^1\,(x^2+y^2)\right)\,\,\,,\,\,\,\,\dot{y}=
\frac{e^U}{\sqrt{2\,y}}\,\left(e_0-\,m^1\,(3\,x^2-y^2)\right)\,,\nonumber\\\dot{x}&=&
2\,\sqrt{2\,y}\,e^U\,\,m^1\,x\,.\label{nks}
 \end{eqnarray}
 The solution in terms of the harmonic functions
 $H_0=h_0+\sqrt{2}\,e_0\,\tau$,  $H^1=h^1-\sqrt{2}\,m^1\,\tau$
 reads \cite{Hotta:2007wz}:
 \begin{eqnarray}
e^{-4\,U}&=&H_0\,(H^1)^3-B^2\,\,\,,\,\,\,\,x=\frac{B}{(H^1)^2}\,\,\,,\,\,\,\,y=\frac{e^{-2\,U}}{(H^1)^2}\,,
 \end{eqnarray}
 where $B$ is a constant defining the value of the axion at
 radial infinity ($\tau=0$).
Using the form of $W_0$ in eq. (\ref{w01}) and eqs. (\ref{nks}) and
(\ref{mmdd}), it is straightforward to compute $\frac{\partial
W_0}{\partial e_0}$ and $\frac{\partial W_0}{\partial m^1}$ and
check, for these components of $\frac{\partial{W_0}}{\partial
\Gamma^M}$, that eq. (\ref{ZdotW})  is satisfied. Analogous
conclusions are obtained for the case $(e_0,m^1)=0$.\par
 As far as
the existence of $\mathcal{W}_3$ is concerned, we can repeat here
the comments made in Sections \ref{sec2} and \ref{sec25}, namely we
can express this prepotential \emph{locally} in terms of the
integral along characteristic lines (which are now geodesics on
${\Scr M}_3$) of the Lagrangian. Of special relevance in this
respect would be manifolds ${\Scr M}_3$ for which the corresponding
$D=3$ Hamiltonian system is \emph{integrable}. In this case a
complete solution $\mathcal{W}_3$, and thus  $\mathcal{W}$, of the
Hamilton--Jacobi equation exists as a function of \emph{globally
defined} integration constants $P_\alpha$ in involution. Candidates
for such integrable models are $D=3$ sigma models based on
\emph{symmetric} target manifolds ${\Scr M}_3$. This  analysis is
still work in progress \cite{wip}.

\section{Conclusions}

In this note we have analyzed the first order description of static
black holes in an Hamiltonian framework, where the prepotential
characterizing the flow has a natural interpretation as Hamilton
principal function. A local form for the solution is available from
the general theory, and was discussed in Section \ref{sec25}, both
for the extremal and non extremal cases. We showed that from
the general form of the solution also follows that the prepotential
is duality invariant.\par
 This kind of analysis, based on the Hamilton--Jacobi
formalism, was fruitfully applied in the context of gauge/gravity
correspondence from the quantum field theory side in a series of
papers pionereed by \cite{de Boer:1999xf}-\cite{Fukuma:2002sb} and
specifically to spherically symmetric, static four dimensional black
holes in \cite{Hotta:2009bm}. From this analysis emerges the role of
the prepotential $\mathcal{W}$, on the dual QFT side, as a
c-function characterizing the renormalization group flow towards the
conformal fixed point. The results in \cite{Hotta:2009bm} apply to
cases where the principal function does not depend explicitly in the
evolution parameter (cut-off) corresponding, on the gravity side, to
extremal solutions. It would be interesting to extended such
analysis also to non-extremal cases, where  no  IR critical point
exists, but nevertheless,the function $\mathcal{W}$ is defined and
should play a role.\par
 It would also
be interesting to extend our approach to more general settings which
include for instance higher derivative corrections to the
supergravity Lagrangian (see \cite{Fukuma:2002sb} and references
therein), stationary (non static) solutions and multicenter black
holes.\par
 Another relevant
issue is related to the duality invariance of the most general
$\mathcal{W}$ function in four dimensional supergravity theories,
which hints to the fact that $\mathcal{W}$ should be expressed only
in terms of duality invariant quantities, constructed in terms of
the central and matter charges.  This was done in
\cite{Andrianopoli:2007gt} for a broad class of extremal black holes
in various supergravities. Such analysis however did not encompass
the five-parameter solutions found in
\cite{Hotta:2007wz,Gimon:2007mh} and  described, for specific
choices of electric and magnetic charges, by the prepotentials given
in \cite{Ceresole:2007wx,LopesCardoso:2007ky}. The prepotentials
found in those papers, however, do not exhibit manifest duality
invariance, being expressed in terms of a reduced set of charges. We
expect that the completion of these prepotentials to a function of a
full set of charges will exhibit manifest duality and would be
expressible in terms of central and matter charges only. Such
completion could be achieved, for instance, by integrating eq.
(\ref{ZdotW}). An equivalent approach to the same problem is to use
duality transformations on the five parameter solution to generate
the remaining charges \cite{marrani2}.\par Finally $D=3$ global
symmetry transformations which map static black holes into one
another, and which are employed in the solution generating
techniques \cite{Cvetic:1995kv,Bergshoeff:2008be}, should be viewed
as canonical transformations in the extended phase space, according
to our discussion, and also along the lines of \cite{canonically}.
We leave to a future investigation the study of the relations
between the various kinds of static black holes within this new
framework. It would be interesting to address, at least for
symmetric scalar manifolds ${\Scr M}_3$, the issue of integrability
of the Hamiltonian system in the extended phase space. To this
respect, of particular relevance, in the case of symmetric
manifolds, is the Lax pair description of the geodesic equations
\cite{Fre:2009et,Chemissany:2009hq,Chemissany:2009af}, and the
existence of a number of conserved Hamiltonians in involution, found
in \cite{Fre:2009dg}.

\section*{Acknowledgements}
Work supported in part by  PRIN
Program 2007 of MIUR.

\end{document}